\documentclass[showpacs,10pt,twocolumn,prl]{revtex4-1}
\usepackage{amsmath}
\usepackage{multirow}
\usepackage{amssymb}
\usepackage{graphics}
\usepackage{epsfig}
\usepackage{color}

\begin{document}

\title{Extreme orbital $ab$-plane upper critical fields far beyond Pauli limit in 4$H_{b}$-Ta(S, Se)$_{2}$ bulk crystals}
\author{Fanyu Meng$^{1,2,\dag}$, Yang Fu$^{1,2,\dag}$, Senyang Pan$^{3}$, Shangjie Tian$^{4,1,2}$, Shaohua Yan$^{1,2}$, Zhengyu Li$^{3}$, Shouguo Wang$^{4}$, Jinglei Zhang$^{3,*}$, and Hechang Lei$^{1,2,*}$}
\affiliation{$^{1}$Department of Physics and Beijing Key Laboratory of Opto-electronic Functional Materials $\&$ Micro-nano Devices, Renmin University of China, Beijing 100872, China\\
$^{2}$Key Laboratory of Quantum State Construction and Manipulation (Ministry of Education), Renmin University of China, Beijing 100872, China\\
$^{3}$Anhui Key Laboratory of Condensed Matter Physics at Extreme Conditions, High Magnetic Field Laboratory, HFIPS, Chinese Academy of Sciences, Hefei 230031, China\\
$^{4}$School of Materials Science and Engineering, Anhui University, Hefei 230601, China\\
}
\date{\today}

\begin{abstract}
Transition metal disulfides 4$H_{b}$-Ta(S, Se)$_{2}$  with natural heterostructure of 1${T}$- and 1${H}$-Ta(S, Se)$_{2}$ layers have became the focus of correlated materials their unique combinations of Mott physics and possible topological superconductivity. In this work, we study the upper critical fields $\mu_{0}H_{c2}$ of 4$H_{b}$-TaS$_{2}$ and 4$H_{b}$-TaS$_{1.99}$Se$_{0.01}$ single crystals systematically. 
Transport measurements up to 35 T show that both of ${ab}$-plane and ${c}$-axis upper critical fields ($\mu_{0}H_{c2,ab}$  and $\mu_{0}H_{c2,c}$) for 4$H_{b}$-TaS$_{2}$ and 4$H_{b}$-TaS$_{1.99}$Se$_{0.01}$ exhibit a linear temperature dependent behavior down to 0.3 K, suggesting the three-dimensional superconductivity with dominant orbital depairing mechanism in bulk 4$H_{b}$-Ta(S, Se)$_{2}$. However, the zero-temperature $\mu_{0}H_{c2,ab}$(0) for both crystals are far beyond the Pauli paramagnetic limit $\mu_{0}H{\rm_{P}}$. 
It could be explained by the effects of spin-momentum locking in 1$H$-Ta(S, Se)$_{2}$ layers with local inversion symmetry broken and the relatively weak intersublattice interaction between 1$H$ layers due to the existence of 1$T$ layers.
\end{abstract}

\maketitle

The exploration of superconductors with large upper critical field $\mu_{0}H_{c2}$ is of great interest to fundamental and applied physics. In conventional superconductors, the application of a magnetic field above the $\mu_{0}H_{c2}$ can destroy superconductivity via orbital or Pauli paramagnetic depairing mechanisms. The former one originates from the interaction between magnetic field and electron momentum. The latter one is caused by spin alignment of Cooper pairs by magnetic field, i.e., the competition between the binding energy of a Cooper pair and the Zeeman splitting energy \cite{4,5}. When the orbital depairing effect is weakened or eliminated, the $\mu_{0}H_{c2}$ is determined mainly by the Pauli paramagnetic effect \cite{6,7}.

However, in some systems, the $\mu_{0}H_{c2}$ of a superconductor can exceed the Pauli paramagnetic limit field $\mu_{0}H_{\rm{P}}$. 
In noncentrosymmetric superconductors, spin-orbit coupling (SOC) lifts the degeneracy of the electron band and manifest as an effective magnetic field $\mu_{0}H_{\rm so}(\boldsymbol{k})$. 
The electron spins are locked along the directions of $\mu_{0}H_{\rm so}(\boldsymbol{k})$ which are opposite for electrons of opposite momenta \cite{8,9,10}. Such spin-momentum locking can significantly enhance the $\mu_{0}H_{c2}$ beyond the $\mu_{0}H_{\rm{P}}$. 
For example, the Rashba-type SOC can lock the spin in the $ab$ plane, which will greatly enhance the $c$-axis upper critical field $\mu_{0}H_{c2,c}$ \cite{11,12}. Another example is Ising superconductors, such as monolayer or few-layer MoS$_2$ \cite{13,14} and NbSe$_2$ \cite{15,16}, in which the $ab$-plane upper critical field $\mu_{0}H_{c2,ab}$ increases far above $\mu_{0}H_{\rm{P}}$ because of the Zeeman-type SOC locks the spin along the ${c}$ axis. 

In centrosymmetric $s$-wave superconductors, such spin-momentum locking is usually destroyed due to the existence of inversion symmetry in the bulk materials and the restored spin degeneracy. However, recent studies have shown that Ising-protected superconductivity can occur in centrosymmetric materials at two-dimensional (2D) limit, such as stanene \cite{17} and PdTe$_{2}$ films \cite{18}, where SOC induces spin-orbit locking near the $\Gamma$ point to increase $\mu_{0}H_{c2,ab}$ \cite{19}. On the other hand, the strong spin-orbital-parity coupling caused by topological band inversion near the topological band crossing can also effectively pin the electron spins and lead to anisotropic renormalization effect of the external Zeeman field, thereby increasing the $\mu_{0}H_{c2,ab}$ anisotropically \cite{20}. Such mechanism has been used to explain the observed large enhancement $\mu_{0}H_{c2,ab}$ of few-layer 2${M}$-WS$_{2}$ \cite{21} and monolayer 1${T}$'-MoTe$_{2}$ \cite{22}.

In contrast, for bulk centrosymmetric superconductors, the enhancement of $\mu_{0}H_{c2}$ is still rare. The enhanced $\mu_{0}H_{c2,ab}$ beyond the $\mu_{0}H_{\rm{P}}$ has been observed in bulk (LaSe)$_{1.14}$(NbSe$_{2}$)$_{m}$ ($m$ = 1, 2) and [(SnSe)$_{1+\delta}$]$_{m}$[NbSe2]$_{1}$ ($m$ = 1 - 15) with misfit structures, and organic cation intercalated bulk NbSe$_{2}$ \cite{23,24,25,26}. However, in these systems, the blocking layers composed of LaSe, SnSe or organic cations can effectively decouple the interlayer coupling between two NbSe$_{2}$ layers. Thus they still show similar features of 2D superconductivity to the monolayer NbSe$_{2}$ with spin-momentum locking, resulting in the large $\mu_{0}H_{c2,ab}$.

In this work, we show that the bulk centrosymmetric superconductors 4$H_{b}$-Ta(S, Se)$_{2}$ exhibit the enhancements of $\mu_{0}H_{c2,ab}$ that are about three times larger than the $\mu_{0}H_{\rm{P}}$ even the dimensionalities of superconductivity are still three dimensional (3D). Such behaviors could originate from the local inversion-symmetry breaking in 1$H$-Ta(S, Se)$_{2}$ layers combined with the weak intersublattice coupling.

The single crystals of $4{H}_{{b}}$-TaS$_{2}$ and $4{H}_{{b}}$-TaS$_{1.99}$Se$_{0.01}$ were grown by using the chemical vapor transport method \cite{28}. 
X-ray diffraction (XRD) patterns were measured using a Bruker D8 X-ray machine with  Cu $K_{\alpha}$ radiation ($\lambda$ = 1.5418 \AA). 
The magnetic susceptibility and transport measurements were measured using the Quantum Design MPMS3 and PPMS-14T. 
High-field transport measurements was performed in Chinese High Magnetic Field Laboratory (CHMFL) in Hefei using a resistive water-cooled magnet in fields up to 35 T and at temperatures down to 0.3 K in a helium-3 cryostat. Field dependence of resistivity was measured by AC bridge (Lakeshore, 370).


\begin{figure}[tbp]
\centerline{\includegraphics[scale=0.17]{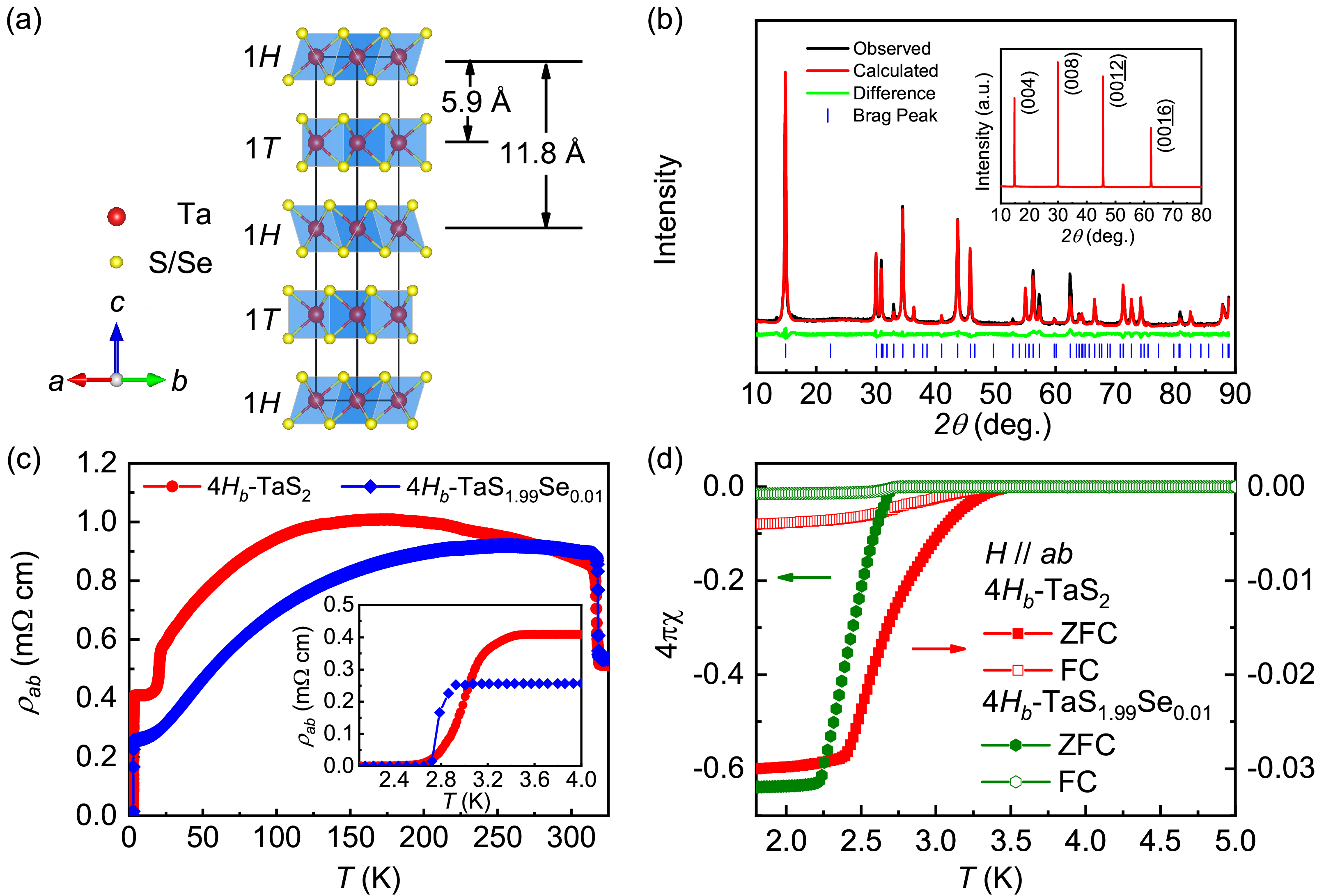}} \vspace*{-0.3cm}
\caption{(a) Crystal structure of $4{H}_{{b}}$-Ta(S, Se)$_{2}$. (b) Powder XRD pattern of crushed $4{H}_{{b}}$-TaS$_{2}$ single crystals. Inset shows the XRD pattern of a  $4{H}_{{b}}$-TaS$_{2}$ single crystal. (c) Temperature dependence of zero-field $\rho_{ab}(T)$ for  $4{H}_{{b}}$-TaS$_{2}$ and $4{H}_{{b}}$-TaS$_{1.99}$Se$_{0.01}$ single crystals. Inset shows the enlarged view of $\rho_{ab}(T)$ curves at low-temperature region. (d) Temperature dependence of 4$\pi\chi(T)$ at $\mu_{0}{H}$ = 1 mT along the ${ab}$ plane with ZFC and FC modes.}
\end{figure}

Ta(S, Se)$_{2}$ has a layered structure with the Ta(S, Se)$_{2}$ layers stacking along $c$ axis with weak van der Waals (vdW) interactions. The Ta(S, Se)$_{2}$ layer can form different local structures and the most typical ones are the ${T}$-type structure with Ta(S, Se)$_{6}$ octahedra layer and the ${H}$-type structure with Ta(S, Se)$_{6}$ trigonal prism layer. 
The $4{H}_{{b}}$-Ta(S, Se)$_{2}$ is one of Ta(S, Se)$_{2}$ polymorphs, which is composed of alternating stackings of 1${H}$- and 1${T}$-Ta(S, Se)$_{2}$ layers (Fig. 1(a)). The interlayer distance $s$ between two 1${H}$-Ta(S, Se)$_{2}$ layers is about 11.8 \AA, when the interlayer distance between 1$H$ and 1$T$ layers is about 5.9 \AA. The $4{H}_{{b}}$-Ta(S, Se)$_{2}$ has the hexagonal symmetry with P6$_{3}$/${mmc}$ space group (No. 194). 
Although the $4{H}_{{b}}$-Ta(S, Se)$_{2}$ crystal has a global inversion symmetry with the inversion center located at the center of the 1${T}$ layer, the 1${H}$ layer has a local inversion symmetry breaking \cite{27}. Fig. 1(b) presents the powder XRD pattern of $4{H}_{{b}}$-Ta(S, Se)$_{2}$ crystal, which can be fitted well by using the crystal structure of $4{H}_{{b}}$-Ta(S, Se)$_{2}$. The inset shows the XRD pattern of a $4{H}_{{b}}$-Ta(S, Se)$_{2}$ single crystal. All of peaks can be indexed by the indices of (00$l$) planes, confirming that the ${c}$ axis is perpendicular to the crystal surface.

Fig. 1(c) shows the temperature dependence of $ab$-plane resistivity  $\rho_{ab}(T)$ of $4{H}_{{b}}$-TaS$_{2}$ and $4{H}_{{b}}$-TaS$_{1.99}$Se$_{0.01}$ single crystals. With the decrease of temperature, the $\rho_{ab}(T)$ curve of $4{H}_{{b}}$-TaS$_{2}$ exhibits two jumps at 315 K and 22 K. The former one is ascribed to the formation of $\sqrt{13}$$\times$$\sqrt{13}$ commensurate charge density wave (CCDW) transition in the 1${T}$ layer and the latter one could be due to the appearance of CCDW in the 1${H}$ layer, both of which are consistent with the results reported previously \cite{28,29,30,31,32}. When lowering temperature further, there is a superconducting transition with the onset transition temperature $T$$_{c}^{\rm{onset}}$ = 3.5 K, which is about 4 times higher than that of 2${H}$-TaS$_{2}$ \cite{33}. It is noted that the transition width $\Delta$$T$ is rather large ($\sim$ 0.9 K) and the zero-resistivity temperature $T$$_{c}^{\rm{zero}}$ is about 2.6 K. In contrast, $4{H}_{{b}}$-TaS$_{1.99}$Se$_{0.01}$ only shows the resistivity jump at $\sim$ 316 K, suggesting that the $\sqrt{13}$$\times$$\sqrt{13}$ CCDW in 1${T}$ layer still exist while the CCDW transition in the 1${H}$ layer is strongly suppressed by just 0.5 \% Se doping. Moreover, the $T$$_{c}^{\rm{onset}}$ of $4{H}_{{b}}$-TaS$_{1.99}$Se$_{0.01}$ (2.92 K) is slightly lower than that of undoped sample but with a narrower $\Delta$$T$ ($\sim$ 0.27 K), in agreement with the results in literature \cite{34}. Fig. 1(d) shows the dc magnetic susceptibility 4$\pi$$\chi$(${T}$) as a function of temperature at $\mu_{0}{H}$ = 1 mT along the ${ab}$ plane with zero-field-cooling (ZFC) and field-cooling (FC) modes. The $T$$_{c}^{\rm{onset}}$ defined as the temperature where the 4$\pi$$\chi$(${T}$) starts to become negative is about 3.47 K for $4{H}_{{b}}$-TaS$_{2}$ and 2.7 K for $4{H}_{{b}}$-TaS$_{1.99}$Se$_{0.01}$, respectively, consistent with the resistivity data. For $4{H}_{b}$-TaS$_{2}$, the ZFC 4$\pi$$\chi$(${T}$) shows that the superconducting volume fraction (SVF) at 1.8 K is only about 3.0 \%, which explains the large $\Delta$$T$ in $\rho$$_{ab}$(${T}$) curve. It implies that the CCDW in 1${H}$ layer may strongly compete with superconductivity, leading to the weak superconducting behavior. In contrast, $4{H}_{{b}}$-TaS$_{1.99}$Se$_{0.01}$ shows a bulk superconductivity with the SVF of about 63.9 \% at 1.8 K when the 1${H}$-layer CCDW is suppressed. Furthermore, the FC curves for both crystals show much small SVFs, implying rather strong flux pinning effects in these type-II superconductors \cite{28}.

\begin{figure}
\centerline{\includegraphics[scale=0.18]{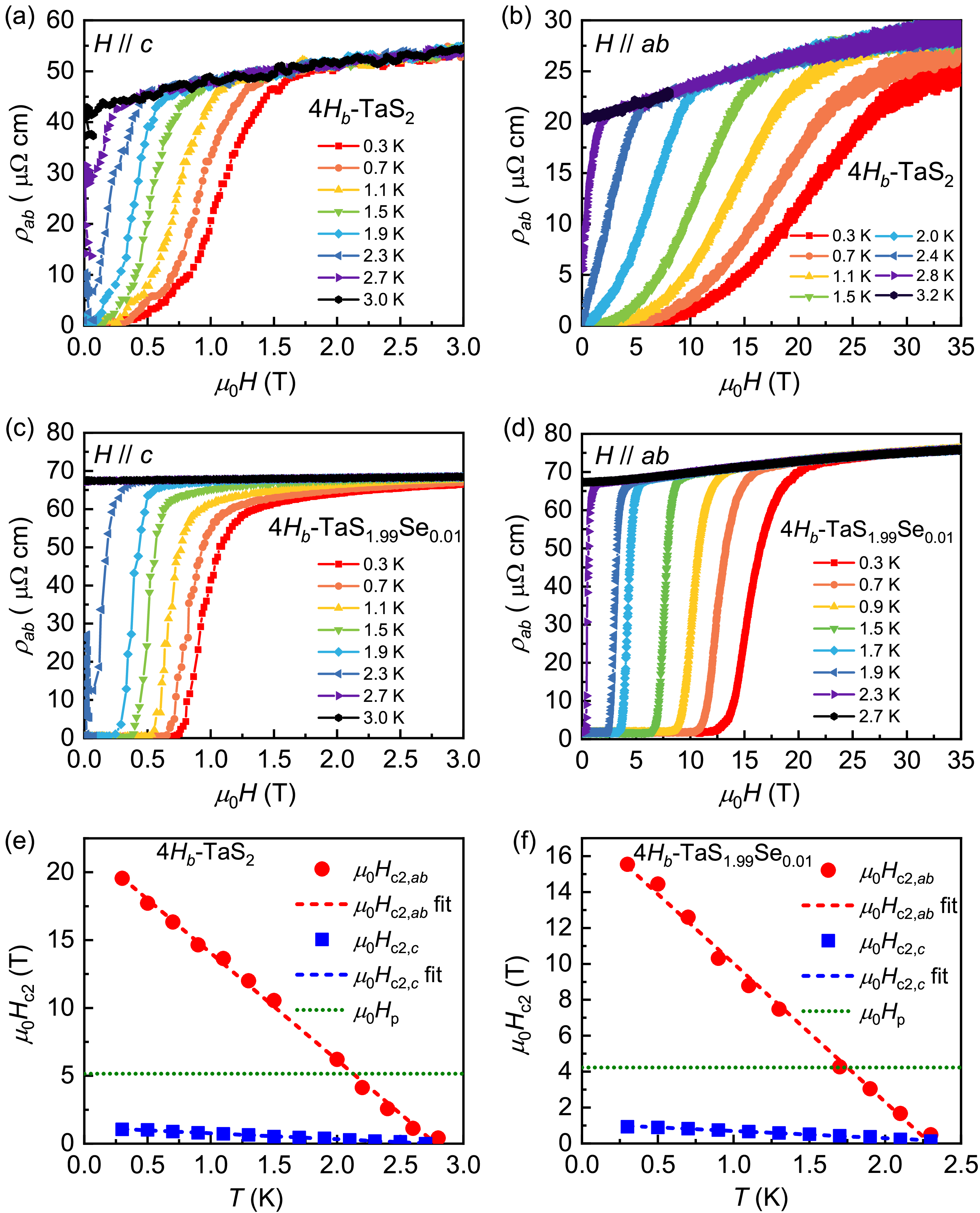}} \vspace*{-0.3cm}
\caption{Field dependence of $\rho_{ab}(\mu_{0}{H})$ of $4{H}_{{b}}$-TaS$_{2}$ single crystal for (a) ${H}\Vert{c}$ and (b) ${H}\Vert{ab}$, and of the $4{H}_{{b}}$-TaS$_{1.99}$Se$_{0.01}$ single crystal for (c) ${H}\Vert{c}$ and (d)${H}\Vert{ab}$ measured at various temperatures in field up to 35 T. (e) and (f) Temperature dependence of $\mu_{0}{H}_{c2}(T)$ for ${H}\Vert{c}$ (red circles) and ${H}\Vert{ab}$ (blue square) for $4{H}_{{b}}$-TaS$_{2}$ and $4{H}_{{b}}$-TaS$_{1.99}$Se$_{0.01}$ single crystal, respectively. Red and blue dashed lines are the fits using 3D GL model for $\mu_{0}{H}_{c2,ab}(T)$ and $\mu_{0}{H}_{c2,c}(T)$. Green dashed lines denote the $\mu_{0}{H}_{\rm{P}}$.}
\end{figure}

Fig. 2 illustrates the resistive transitions from superconducting state to normal state with an applied magnetic field up to 35 T which oriented parallel to the ${c}$ axis and parallel to the ${ab}$ plane for both crystals. It can be clearly seen that superconductivity is suppressed by increasing magnetic field at the same temperature.
At $T$ = 0.3 K, the superconductivity of $4{H}_{{b}}$-TaS$_{2}$ is completely suppressed at 1.6 T for  ${H}\Vert{c}$ (Fig. 2(a)). In contrast, this field is significantly enhanced to about 27 T for ${H}\vert\vert{ab}$ (Fig. 2(b)). In addition, for both field directions, the superconducting transitions of $\rho_{ab}({T})$ curves for $4{H}_{{b}}$-TaS$_{2}$ are shifted to lower magnetic fields gradually with increasing temperatures. 
For $4{H}_{{b}}$-TaS$_{1.99}$Se$_{0.01}$, the $\rho_{ab}(\mu_{0}{H})$ curves exhibit similar behaviors to those of $4{H}_{{b}}$-TaS$_{2}$ and the superconductivity is suppressed at ${\mu_{0}{H}}\sim$ 1.25 T for ${H}\Vert{c}$ (Fig. 2(c)) and at $\mu_{0}{H}\sim$ 20 T for ${H}\Vert{ab}$ (Fig. 2(d)), respectively. It is worthy of noting that the superconducting transition widths of $4{H}_{{b}}$-TaS$_{1.99}$Se$_{0.01}$ is much narrower than those of $4{H}_{{b}}$-TaS$_{2}$ at various temperatures, possibly related to the bulk superconductivity in the former.

Figs. 2(e) and 2(f) show the $\mu_{0}{H}_{c2}({T})$ for ${H}\Vert{c}$ and ${H}\Vert{ab}$ as a function of temperature for $4{H}_{{b}}$-TaS$_{2}$ and $4{H}_{{b}}$-TaS$_{1.99}$Se$_{0.01}$ single crystal, respectively. Because the $4{H}_{{b}}$-TaS$_{2}$ has a relatively large $\Delta$$T$, the $\mu_{0}{H}_{{c2}}({T})$ is evaluated using the criterion of 50 \% normal-state resistivity $\rho_{n,ab} (\mu_{0}{H},{T})$. The $\rho_{n,ab} (\mu_{0}{H},{T})$ was determined by linearly extrapolating the normal-state behavior above the onset of superconductivity transition in $\rho_{ab}(\mu_{0}{H})$ curves. Interestingly, the $\mu_{0}{H}_{{c2}}({T})$ curves of $4{H}_{{b}}$-TaS$_{2}$ and $4{H}_{{b}}$-TaS$_{1.99}$Se$_{0.01}$ single crystals exhibit similar linear behaviors in the whole temperature range for both field directions. We fitted these linear $\mu_{0}{H}_{{c2}}$-${T}$ relationships for both $ab$-plane and $c$-axis fields in the framework of a phenomenological 3D anisotropic Ginzburg-Landau (GL) theory with considering orbital depairing mechanism only. In general, orbital depairing occurs when the vortices begin to overlap at the orbital critical field $\mu_{0}{H}_{{c2}}^{\rm{orb}} \sim \frac{\Phi_{0}}{\xi_{0}}$ with $\Phi_{0}$ being magnetic flux quantum (= 2.07×10$^{-15}$ Wb) and $\xi$ being the GL coherence length. In highly anisotropic materials, the GL coherence length may vary in different directions of the material. Due to the layered structure of $4{H}_{{b}}$-Ta(S, Se)$_{2}$, the GL coherence length  $\xi_{ab}$ in the ${ab}$-plane can be assumed isotropic, which may distinctly different from the ${c}$-axis one. In the presence of a ${c}$-axis magnetic field, the $\mu_{0}{H}_{{c2,c}}({T})$ depends only on $\xi_{ab}$ \cite{35,36,37} and $\mu_0 H_{c 2, c}(T)=\frac{\Phi_0}{2 \pi \xi_{a b}(T)^2}=\frac{\Phi_0}{2 \pi \xi_{a b}(0)^2}\left(1-\frac{T}{T_c}\right)$, where $\mu_{0}$ is vacuum permeability and the temperature-dependent $\xi_{a b}(T)=\xi(0)\left(1-\frac{T}{T_c}\right)^{-\frac{1}{2}}$. While the $\mu_0 H_{c 2, ab}(T)$ for ${H}\vert\vert{ab}$ depends on both $\xi_{ab}$ and $\xi_{c}$ with $\mu_0 H_{c 2, ab}(T) =\frac{\Phi_0}{2 \pi \xi_{ab}(T) \xi_c(T)}=\frac{\Phi_0}{2 \pi \xi_{ab}(0) \xi_c(0)}\left(1-\frac{T}{T_c}\right)$. Thus, for the orbital depairing mechanism, the $\mu_0 H_{c 2,c}(T)$ and $\mu_0 H_{c 2,ab}(T)$ should exhibit a linear temperature dependence. It is clearly seen that the 3D GL equations can fit the $\mu_0 H_{c 2,c}(T)$ and $\mu_0 H_{c 2,ab}(T)$ curves for both crystals perfectly. The fitted zero-temperature $\mu_0 H_{c 2, c}(0)$ and $\mu_0 H_{c 2,ab}(0)$ of $4{H}_{{b}}$-TaS$_{2}$ is 1.20(1) T and 22.0(3) T, respectively (Fig. 2(e)). Correspondingly, the calculated $\xi_{a b}(0)=\sqrt{\frac{\Phi_0}{2 \pi \mu_0 H_{c 2, c}(0)}} = 165.4(7)$ \AA{\ }and $\xi_{c}(0) = 9.1(5)$ \AA. For $4{H}_{{b}}$-TaS$_{1.99}$Se$_{0.01}$, $\mu_0 H_{c 2, c}(0)$ = 1.10(2) T and $\mu_0 H_{c 2,ab}(0)$ = 17.7(3) T (Fig. 2(f)). The calculated $\xi_{a b}(0)$ is 173(2) \AA{\ }when $\xi_{c}(0) = 10.6(3)$ \AA.

For the layered materials, the $\mu_{0}{H}_{{c2,ab}}({T})$ usually has  a linear behavior near $T{_c}$ because when the $\xi_c(T)$ is larger than the interlayer distance $s$ the system behaves like a 3D system and the $\mu_{0}{H}_{{c2,ab}}({T})$ can be described by the 3D GL equation \cite{35}. However, according to the Lawrence-Doniach (LD) model for layered superconductors with weak interlayer Josephson coupling, the temperature dependence of $\mu_{0}{H}_{{c2}}$ can change to $\left(T_{\mathrm{c}}-T\right)^{1 / 2}$ when the $\xi_c(T)$ decreases with temperature and the criterion for the crossover of $\mu_{0}{H}_{{c2}}$ from 3D behavior to 2D one is $\xi_{\mathrm{c}} / \frac{s}{\sqrt{2}}<1$ \cite{38}. Such crossover behavior has been observed in the artificial multilayers with increasing the thickness of nonsuperconducting layer \cite{39}. For present two materials, the values of $\xi_{\mathrm{c}} / \frac{s}{\sqrt{2}}$ are about 1.09 $\sim$ 1.27, which is large than 1, thus both of them should still be the 3D superconducting systems and this explains the linear behavior persists to the temperature far below $T{_c}$.

The most striking feature of $\mu_{0}{H}_{{c2,ab}}({T})$ is that such a linear behavior can be extended far beyond the $\mu_{0}{H}_{\rm{P}}$. For $\mu_{0}{H}_{\rm{P}}$, once the magnetic energy is of the order of the superconducting condensation energy, the system gains energy entering the normal state, thus leading to the $\mu_0 H_{\mathrm{P}} \sim k_{\mathrm{B}} T_{\mathrm{c}} / \sqrt{\chi_{\mathrm{n}}-\chi_{\mathrm{sc}}(T)}$, where $\chi_{\mathrm{n}}$ and ${\chi_{\mathrm{sc}}}$ is magnetic susceptibility at normal state and superconducting state, respectively \cite{40}. For weakly coupled Bardeen-Cooper-Schrieffer (BCS) superconductors with a pure Pauli susceptibility, $\mu_0 H_{\mathrm{P}}=\frac{\Delta_0}{\sqrt{2} \mu_B} \approx 1.86 T_{\mathrm{c}}$, where $\mu_B$ is Bohr magneton and $\Delta_0$ is the superconducting energy gap based on BCS theory for $T$ = 0 K, which is known as the Clogston-Chandrasekhar limit \cite{4,5}. If the BCS Pauli paramagnetic effect is the strong depairing mechanism and the $\mu_{0}{H}_{{c2}}$ should be mainly limited by the $\mu_{0}{H}_{\rm{P}}$. But it can be seen that both $\mu_{0}{H}_{{c2,ab}}({0})$ of $4{H}_{{b}}$-TaS$_{2}$ and $4{H}_{{b}}$-TaS$_{1.99}$Se$_{0.01}$ crystals are about three times larger than their $\mu_{0}{H}_{\rm{P}}$.

The strong scattering due to SOC could lead to the enhancement of $\mu_{0}{H}_{{c2}}$ because of the weakening influence of spin paramagnetism \cite{3,41}. But this theory is valid only for superconductors in the dirty-limit, i.e., $l<\xi$, where $l$ is the mean free length. Based on the transport measurements at normal state \cite{28}, the estimated $l$ of electrons and holes from carrier mobilities and densities are between 941 \AA\ $\sim$ 1570 \AA, much larger than  $\xi_{ab}$(0) and  $\xi_{c}$(0) for both crystals. Thus, the superconductivity of $4{H}_{{b}}$-Ta(S, Se)$_{2}$ should be in the clean-limit and the effect of scattering due to SOC could not interpret their enhanced $\mu_{0}{H}_{{c2,ab}}$.
On the other hand, such remarkable enhancements of $\mu_{0}{H}_{{c2,ab}}({0})$ beyond $\mu_{0}{H}_{\rm{P}}$ are very similar to the phenomena observed in Ising superconductors with spin-momentum locking, such as few-layer or monolayer $2{H}$-MoS$_2$, $2{H}$-NbSe$_2$ and $2{H}$-TaS$_2$ \cite{13,15,42}. 
However, such spin-momentum locking would be destroyed in the bulk crystals where inversion symmetry and spin degeneracy are restored. For example, the $\mu_{0}{H}_{{c2,ab}}({0})$ of bulk 2$H$-TaS$_2$ with $T_c$ = 1.4 K is only about 1.4 T, much smaller than $\mu_{0}{H}_{\rm{P}}$ $\sim$ 2.6 T \cite{43}. 
In contrast, although the inversion symmetry of $4{H}_{{b}}$-Ta(S, Se)$_{2}$ crystals is maintained in the bulk material, their crystal structure comprises two sublattices (1$T$ and 1$H$ layers) and the 1$H$ sublattice lacks inversion symmetry, leading to the local inversion-symmetry breaking. Because of the existence of 1$T$ layers, two 1$H$ layers related by inversion symmetry are only weakly coupled and $4{H}_{{b}}$-Ta(S, Se)$_{2}$ can be regarded as the two copies of a noncentrosymmetric $1{H}$-Ta(S, Se)$_{2}$ with weak interlayer coupling. 
Therefore, the spin-momentum locking will still manifest their effect in bulk $4{H}_{{b}}$-Ta(S, Se)$_{2}$, resulting in the $\mu_{0}{H}_{{c2,ab}}(0)$ far above the Clogston-Chandrasekhar limit $\mu_{0}{H}_{\rm{P}}$. 
More specifically, in noncentrosymmetric materials, the SOC due to the local lack of inversion symmetry has an important effect on the $\mu_{0}{H}_{{\rm{P}}}$ by changing the spin susceptibility [40]. For example, if the intersublattice couplings between two $1{H}$-Ta(S, Se)$_{2}$ layers are zero and the field is chosen to be perpendicular to the SOC, here ${H}\Vert{ab}$, then an external field will have no effect, i.e., the $\chi_{\mathrm{n}}$ will decrease to zero, and the $\mu_{0}{H}_{\rm{P}}$ will diverge, as long as the Zeeman energy is much less than any interband separation energy \cite{40}. In this case, the $\mu_{0}{H}_{{c2,ab}}$ will be determined by the orbital depairing mechanism only and the GL equation will be valid. 
It is noted that the above discussion on the enhancement of $\mu_{0}{H}_{{c2,ab}}$ in $4{H}_{{b}}$-Ta(S, Se)$_{2}$ is qualitative. Further experimental and theoretical studies are needed to fully understand the effects of intersublattice couplings and spin-moment locking on $\mu_{0}{H}_{{c2,ab}}$. In addition, when compared with $4{H}_{{b}}$-TaS$_{1.99}$Se$_{0.01}$, the slight enhancement of the $\mu_{0}{H}_{{c2,ab}}$ in  $4{H}_{{b}}$-TaS$_{2}$ may be due to its increased $T_c$.

\begin{figure}
\centerline{\includegraphics[scale=0.18]{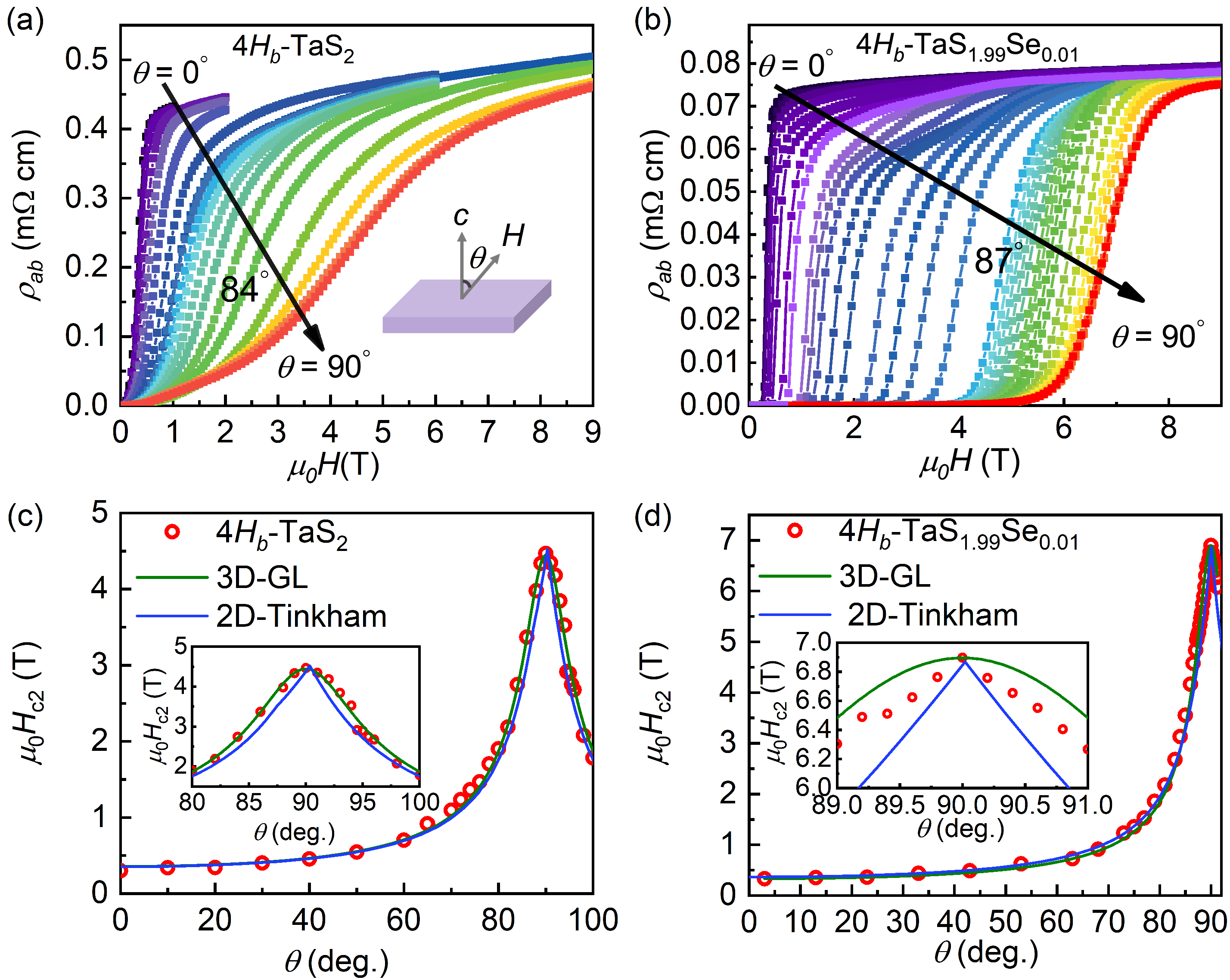}} \vspace*{-0.3cm}
\caption{Field dependence of $\rho_{ab}(\mu_{0}{H})$ at 2 K and various field directions for (a) 4$H_{b}$-TaS$_{2}$ and (b) 4$H_{b}$-TaS$_{1.99}$Se$_{0.01}$ single crystals. Angular dependence of $\mu_{0}{H}_{{c2}}(\theta)$  of (c) 4$H_{b}$-TaS$_{2}$ and (d) 4$H_{b}$-TaS$_{1.99}$Se$_{0.01}$ single crystals at 2.0 K. The $\mu_{0}{H}_{{c2}}(\theta)$ at each $\theta$ are determined using the criterion of 50 \% normal-state resistivity $\rho_{n,ab}(\mu_{0}{H},{\theta})$. The green and blue curves in (c) and (d) represent the fits using 3D anisotropic GL and 2D Tinkham models. Insets: enlarged views of $\mu_{0}{H}_{{c2}}(\theta)$ near $\theta=90^{\circ}$ (${H}\vert\vert{ab}$).}
\end{figure}

To further investigate the dimensionality of superconductivity in $4{H}_{{b}}$-Ta(S, Se)$_{2}$, we studied the angle dependence of the $\mu_{0}{H}_{{c2}}(\theta)$ at 2 K, where $\theta$ is the angle between the magnetic field and the ${c}$ axis of the crystal. Figs. 3(a) and 3(b) show the evolution of $\rho_{ab}(\mu_{0}{H})$ as a function of field at different $\theta$ for 4$H_{b}$-TaS$_{2}$ and 4$H_{b}$-TaS$_{1.99}$Se$_{0.01}$. When $\theta=0^{\circ}$, the superconductivity is suppressed at a relatively low field. With increasing $\theta$, the superconducting transition shifts to higher fields gradually, but when $\theta$ is close to $90^{\circ}$ this shift becomes much faster than those at low-angle region and the $\mu_{0}{H}_{{c2}}$ reaches the maximum value at $\theta=90^{\circ}$. These results further confirm the strong anisotropy of superconductivity in 4$H_{b}$-Ta(S, Se)$_{2}$. For 3D interlayer Josephson-coupled superconductors, the angular dependence of $\mu_{0}{H}_{{c2}}(\theta)$ can be described by the anisotropic 3D GL model \cite{35}, $\left(\frac{\mu_0 H_{c 2}(\theta) \cos (\theta)}{\mu_0 H_{c 2, c}}\right)^2+\left(\frac{\mu_0 H_{c 2}(\theta) \sin (\theta)}{\mu_0 H_{c2, a b}}\right)^2=1$. The general feature of this model is that the $\mu_{0}{H}_{{c2,ab}}(\theta)$ curve is smooth and has a bell shape near $\theta=90^{\circ}$. In contrast, the Tinkham model is used to express the $\mu_{0}{H}_{{c2,ab}}(\theta)$ of 2D superconductors with decoupled interlayer interactions \cite{44}, $\left|\frac{\mu_0 H_{c 2}(\theta) \cos (\theta)}{\mu_0 H_{c 2, c}}\right|+\left(\frac{\mu_0 H_{c 2}(\theta) \sin (\theta)}{\mu_0 H_{c 2, a b}}\right)^2=1$. This equation exhibits a finite slope at $\theta=90^{\circ}$, making a cusp. Figs. 3(c) and 3(d) show the angular dependence of $\mu_{0}{H}_{{c2}}(\theta)$ at 2 K for $4{H}_{{b}}$-TaS$_{2}$ and $4{H}_{{b}}$-TaS$_{1.99}$Se$_{0.01}$ single crystals extracted from Fig. 3(a) and 3(b) using the criterion of 50 \% normal-state resistivity $\rho_{ab}(\mu_{0}{H},{\theta})$. For $4{H}_{{b}}$-TaS$_{2}$, it can be seen that both 3D anisotropic GL model (green) and 2D Tinkham model (blue) can fit the data at low-angle region $\left(\theta \leq 80^{\circ}\right)$ well, whereas for $\theta>80^{\circ}$(inset of Fig. 3(c)), the $\mu_{0}{H}_{{c2}}(\theta)$ curve with rounded bell shape suggests that the 3D GL model can described the behavior of $\mu_{0}{H}_{{c2}}(\theta)$ better and thus $4{H}_{{b}}$-TaS$_{2}$ should be a 3D superconducting system, which is consistent with above analysis of temperature dependence of $\mu_{0}{H}_{{c2,ab}}({T})$. On the other hand, for 4$H_{b}$-TaS$_{1.99}$Se$_{0.01}$, the 3D GL model can fit the $\mu_{0}{H}_{{c2}}(\theta)$ curve better until $\theta$ is very close to $90^{\circ}$$\left(\left|90^{\circ}-\theta\right|< \pm 0.5^{\circ}\right)$, where the curve rises sharply and results in a cusp (inset of Fig. 3(d)). Similar behavior has been observed in Bi$_2$Sr$_2$CaCu$_2$O$_8$ thin films \cite{45} and Nb/CuMn multilayer system \cite{39}. Especially, for Nb/CuMn multilayer, when the cusp-like behavior at $\theta=0^{\circ}$ appears on the top of a bell-shaped curve at 4.2 K, the $\mu_{0}{H}_{{c2,ab}}({T})$ still shows a linear behavior at this temperature region \cite{39}. Such behaviors may be explained by the system near the point of 2D - 3D crossover of superconductivity \cite{46} because $4{H}_{{b}}$-Ta(S, Se)$_{2}$ have the values of $\xi_{\mathrm{c}} / \frac{s}{\sqrt{2}}$ just slightly larger than 1.

In summary, we investigated the superconducting properties of 4$H_{b}$-TaS$_{2}$ and 4$H_{b}$-TaS$_{1.99}$Se$_{0.01}$ single crystals. 
Both of them show the linear temperature dependence of $\mu_{0}H_{c2,ab}(T)$ for ${H}\Vert{ab}$ and ${H}\Vert{c}$, suggesting the 3D superconductivity of 4$H_{b}$-Ta(S, Se)$_{2}$. It is confirmed further by the measurements of angle dependence of $\mu_{0}H_{c2}(\theta)$. Peculiarly, even the 3D orbital depairing mechanism effect is dominant in centrosymmetric 4$H_{b}$-Ta(S, Se)$_{2}$ bulk crystals, they still exhibit rather high $\mu_{0}H_{c2,ab}$(0) ($\sim$ 22 T for 4$H_{b}$-TaS$_{2}$ and $\sim$ 18 T for 4$H_{b}$-TaS$_{1.99}$Se$_{0.01}$), which are about 4 times as large as the $\mu_{0}H_{\rm{P}}$. Such phenomena can be explained by the Ising-pairing enhanced $\mu_{0}H_{c2,ab}$, which is closely related to the heterostructure of bulk 4$H_{b}$-Ta(S, Se)$_{2}$ with local inversion-symmetry breaking of 1$H$ layers and the weak intersublattice interaction of 1$H$ layers because of the existence of 1$T$ layers. Therefore, 4$H_{b}$-Ta(S,Se)$_{2}$ system provides a paradigm that an extreme large $\mu_{0}H_{c2,ab}$ far beyond $\mu_{0}H_{\rm{P}}$ can still be realized in the 3D bulk superconductors with unique local structural symmetry.

This work was supported by Beijing Natural Science Foundation (Grant No. Z200005), National Key R\&D Program of China (Grants Nos. 2022YFA1403800), National Natural Science Foundation of China (Grants No. 12274459, 12241406), the Fundamental Research Funds for the Central Universities and Research Funds of Renmin University of China (RUC) (Grants Nos. 18XNLG14, 19XNLG17 and 21XNLG26), the Outstanding Innovative Talents Cultivation Funded Programs 2022 of Renmin University of China, Beijing National Laboratory for Condensed Matter Physics, and Collaborative Research Project of Laboratory for Materials and Structures, Institute of Innovative Research, Tokyo Institute of Technology.

$^{\dag}$ F.Y.M and Y.F. contributed equally to this work.

$\ast$ Corresponding authors: zhangjinglei@hmfl.ac.cn (J. L. Z.), hlei@ruc.edu.cn (H. C. L.).

\end{document}